\documentclass{article}

\usepackage{arxiv}

\usepackage[utf8]{inputenc} 
\usepackage[T1]{fontenc}    
\usepackage[colorlinks = true,
            linkcolor = blue,
            urlcolor  = blue,
            citecolor = blue,
            anchorcolor = blue]{hyperref}       
\usepackage{url}            
\usepackage{booktabs}       
\usepackage{amsfonts}       
\usepackage{nicefrac}       
\usepackage{microtype}      
\newcommand{\ora}[1]{\ensuremath{\overrightarrow{#1}}} 
\newcommand{\norm}[1]{\left\lVert#1\right\rVert}

\usepackage{caption}
\usepackage{subcaption}
\usepackage{verbatim}
\usepackage{float}
\usepackage{tikz}
\usepackage{newfloat}
\DeclareFloatingEnvironment[name={Supplementary Figure}]{suppfigure}

\usepackage{cleveref}

\usepackage{amsmath,amsfonts,amssymb}

\title{Sensory Regimes of Effective Distributed Searching without Leaders}

\author{
 Ravid Cohen \\
  School of Computer Science,\\
  Tel-Aviv University, Tel-Aviv 69978, Israel
  \And
  Yossi Yovel \\
  Sagol School of Neuroscience,\\ 
  Tel Aviv University, Tel Aviv 39040, Israel
  \And
  Dan Halperin \\
  School of Computer Science,\\
  Tel-Aviv University, Tel-Aviv 69978, Israel
}

\begin{document}
\maketitle

\begin{abstract}
Collective animal movement fascinates children and scientists alike. One of the most commonly given explanations for collective animal movement is improved foraging. Animals are hypothesized to gain from searching for food in groups.  Here, we use a computer simulation to analyze how moving in a group assists searching for food. We use a well-established collective movement model that only assumes local interactions between individuals without any leadership, in order to examine the benefits of group searching. We focus on how the sensory abilities of the simulated individuals, and specifically their ability to detect food and to follow neighbours, influence searching dynamics and searching performance. We show that local interactions between neighbors are sufficient for the formation of groups, which search more efficiently than independently moving individuals. Once a member of a group finds food, this information diffuses through the group and results in a convergence of up to 85\% of group members on the food. Interestingly, this convergence behavior can emerge from the local interactions between group members without a need to explicitly define it. In order to understand the principles underlying the group's performance, we perturb many of the model's basic parameters, including its social, environmental and sensory parameters. We test a wide range of biological-plausible sensory regimes, relevant to different species and different sensory modalities and examine how they effect group-foraging performance. This thorough analysis of  model parameters allows for the generalization of our results to a wide range of organisms, which rely on different sensory modalities, explaining why they move and forage in groups.
\end{abstract}


\section*{Introduction} Why do so many animals move in groups? One of the most accepted benefits of moving in a group is hypothesized to be improved foraging, which is achieved via enhanced sensing. Specifically, searching for food in a group, has been hypothesized to improve searching by allowing individuals to glean information from conspecifics \cite{Berdahl2013b, Creel2014, Cvikel2015b, Gordon2014, Danchin2004}. Group foraging has been suggested to be especially important when searching for an ephemeral resource. This has also been backed by several mathematical frameworks \cite{Clark1986, Pulliam1982, Karpas2017}. In parallel to these analytic models of group foraging, many models,  mostly numeric (but not exclusively), try to explain how animals maintain a group while moving \cite{Attanasi2014, Couzin2005, Gautrais2012, Nagy2010, Ouellette2015, Parrish1999, Sumpter2008, Bialek2012}. Most of these models assume that collective movement emerges from the behavior of individuals and their local interactions with neighbors (sometimes including distant neighbors). The group does not need leaders, and there is no necessity for individual recognition, or signaling to achieve such coordinated behavior. However, only a few attempts have been made to use such collective movement simulations to test the improved searching hypothesis, that is to use a movement model to simulate collective movement and to examine if and how the groups that are formed by this model benefit from enhanced sensing while searching for food. Two examples include \cite{Lihoreau2017, Torney2011}) but they mostly focus on how group foraging improves tracking of the food landscape. Several other models that suggest how group searching may assist sensing (e.g \cite{Berdahl2013b, Suzuki:1996:ACX:647383.724817})  focus on the situation where the target generates a (noisy) gradient that can be sensed and utilized from a distance. This scenario is relevant for olfactory based searching or for light and shade patterns, but with most sensory modalities (e.g., vision, echolocation, tactile sensing) searching is a binary problem with food being either detectable or not. Some additional attempts to model collective searching have been made in collective robotics (e.g., \cite{Suzuki:1996:ACX:647383.724817, Agmon2008, Chan2006, Nagavalli2014}) but these models often rely on non-biological assumptions, such as active transmission of location-information between individuals (agents) or division of labor (but see \cite{Giuggioli2016FromAT} for a biological approach). 

In this work, we use a simple well-studied movement model \cite{Couzin2002} to examine how moving in a group improves searching. We focus on the sensory aspects of a commonly used agent-based movement model. We test a wide range of biological-plausible sensing regimes, relevant to different species relying on different sensory modalities and we examine how they effect group-foraging performance. We consider a scenario of $n$ agents moving in a two-dimensional region while searching for a sparse resource. To simplify the problem, we simulated a single target (representing food) with an unknown location at each simulation. In the simulations, all agents start moving from the center of the region, each in a random direction. At every time step, each agent adjusts its movement according to its current direction and according to the agents around it (see below). The agents’ steps are thus characterized by a constant step-length (i.e., speed is constant) and a continuously changing direction. To simplify the sensory model, we assumed that the agents sense equally in all directions, and can be sensed from all directions. We define two sensing radii: $r^t$ is the detection radius of the target, and $r^s$ is the detection radius of neighboring individuals. In the case of visual animals, these radii represent the visual range for detecting food and a conspecific respectively, and in the case of echolocating bats or dolphins, where sensing is based on sound, they would represent the echo detection range of a food item, and the eavesdropping range on conspecifics respectively. In reality, the values of these two radii will depend on the sensory modality and the characteristics of the system (e.g., the size of the food and the behavior of the conspecific). We tested several combinations of biologically-reasonable radii and analyzed how they influence group foraging.

At each time step an agent advances according to its velocity and direction,
\begin{align}
	p^{t+1}_i = p^t_i+xd^t_i,
\end{align}
where $p^t_i$ is the location of agent $i$ at time $t$, $x$ is the agent's step magnitude, and $d^t_i$ is the unit direction vector of an agent $i$ at time $t$. The agent's step magnitude is defined by,
\begin{align}
	x = \frac{v}{c_f},
\end{align}
where $v$ is the agent's velocity and $c_f$ is the sensory direction update rate, i.e., how often is sensory information collected and the movement updated accordingly. This parameter represents an actual biological feature that might be adapted through evolution and we therefore examined how it influences group foraging performance.

While searching, the agent's movement direction is determined based on its previous direction and on the location and movement direction of its neighbors. When moving socially, the agent is influenced by neighboring agents as follows: 
\begin{align}\label{bat direction}
	\ora{d^i} = \frac{(1-\rho)\ora{d^i_{\textrm{self}}}+\rho\ora{d^i_{\textrm{social}}}}{\norm{(1-\rho)\ora{d^i_{\textrm{self}}}+\rho\ora{d^i_{\textrm{social}}}}_2},
\end{align}
where $\ora{d^i}$, as already mentioned, is the direction of an agent $i$, $\ora{d^i_{\textrm{self}}}$ and $\ora{d^i_{\textrm{social}}}$ are the individual and social effects on agent $i$ respectively. The parameter $\rho$ determines the relative weight of the self and the social effects. Note that when $\rho$ is set to zero, the agents are not affected by their neighbors and move independently.

When moving as independent individuals, each of our agents applied a correlated walk with Gaussian noise represented by a distribution of turning angles whose width ($\sigma$) we manipulated. The individual movement direction, $\ora{d^i_{\textrm{self}}}$, can be described as follow,
\begin{gather}
		d^{i(t)}_{\textrm{self}} = 
	\begin{cases}
		\alpha, & t=0\\
		d^{i(t-1)} + \beta & t>0,
	\end{cases}\\
	\alpha \sim \mathcal{U}[0,360),\\
	\beta \sim \mathcal{N}(0,\sigma).
\end{gather}
where $d_{\textrm{self}}^{i}$ and $d^{i}$ are the angles (in degrees) that the vector $\ora{d_{\textrm{self}}^{i}}$ and $\ora{d^{i}}$ form with the positive x-axis respectively, and $\sigma$ represents the width of the distribution of turning angles. At the first step, $d_{\textrm{self}}^{i}$ is drawn from a uniform angle distribution and afterwards it is determined according to the direction of the agent in the previous step with the addition of Gaussian noise. 

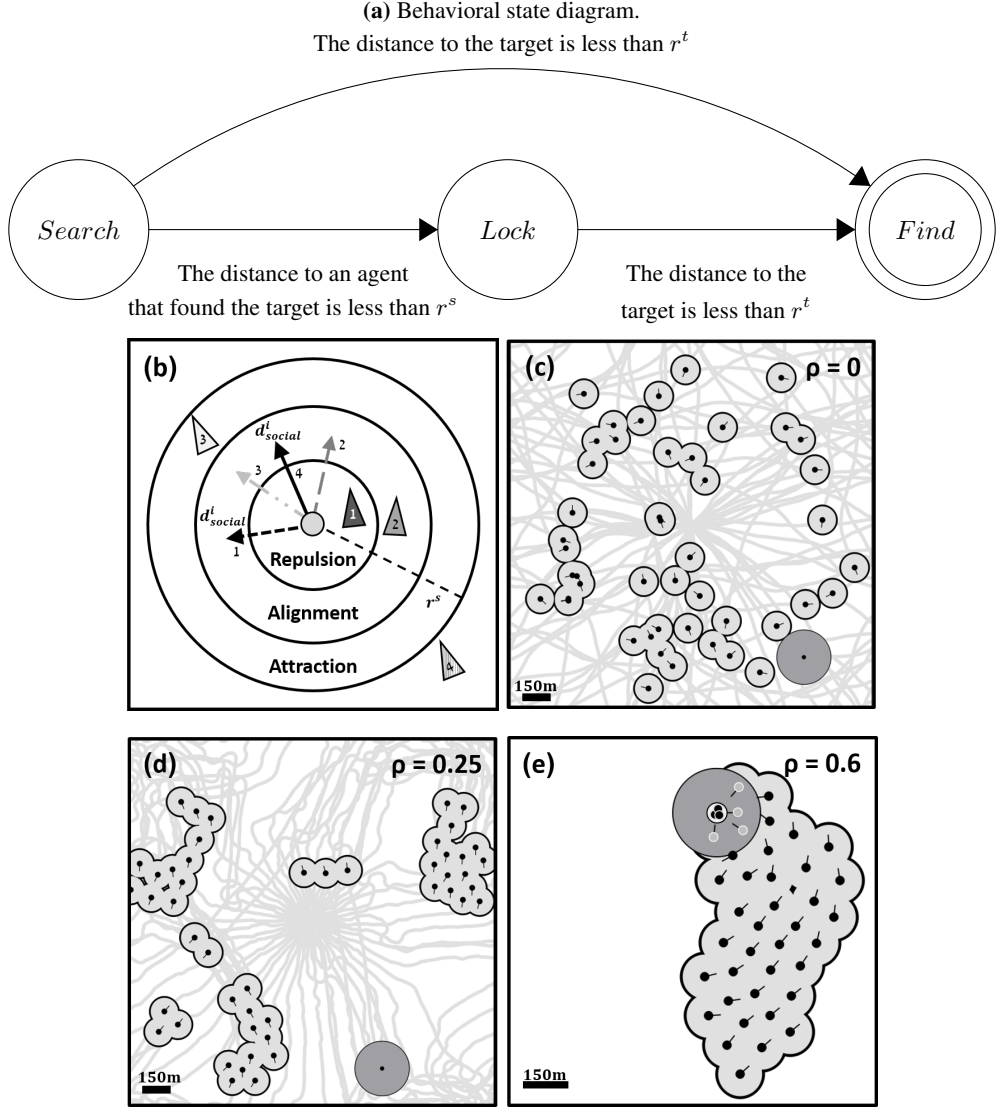
\begin{figure*}
	\centering
	\begin{subfigure}{1\textwidth}
    \captionsetup{format=plain, labelfont=bf}
	\caption{Behavioral state diagram.}
	\vspace{-1em}
	\begin{center}
	\begin{tikzpicture}[scale=0.31]
		\tikzstyle{every node}+=[inner sep=0pt]
		\draw [black] (18.1,-24.3) circle (3);
		\draw (18.1,-24.3) node {$Search$};
		\draw [black] (54.4,-24.3) circle (3);
		\draw (54.4,-24.3) node {$Find$};
		\draw [black] (54.4,-24.3) circle (2.4);
		\draw [black] (36.5,-24.3) circle (3);
		\draw (36.5,-24.3) node {$Lock$};
		\draw [black] (21.1,-24.3) -- (33.5,-24.3);
		\fill [black] (33.5,-24.3) -- (32.7,-23.8) -- (32.7,-24.8);
		\draw (27.3,-25.8) node [below] {\small The distance to an agent};
		\draw (27.3,-27.2) node [below] {\small that found the target is less than $r^s$};
		\draw [black] (39.5,-24.3) -- (51.4,-24.3);
		\fill [black] (51.4,-24.3) -- (50.6,-23.8) -- (50.6,-24.8);
		\draw (45.45,-25.8) node [below] {\small The distance to the};
		\draw (45.45,-27.2) node [below] {\small target is less than $r^t$};
		\draw [black] (20.446,-22.432) arc (125.38279:54.61721:27.294);
		\fill [black] (52.05,-22.43) -- (51.69,-21.56) -- (51.11,-22.38);
		\draw (36.25,-16.89) node [above] {\small The distance to the target is less than $r^t$};
	\end{tikzpicture}
	\end{center}
    \label{1a}
	\end{subfigure}
	\begin{subfigure}{0.30\textwidth}
		\includegraphics[width=1\textwidth]{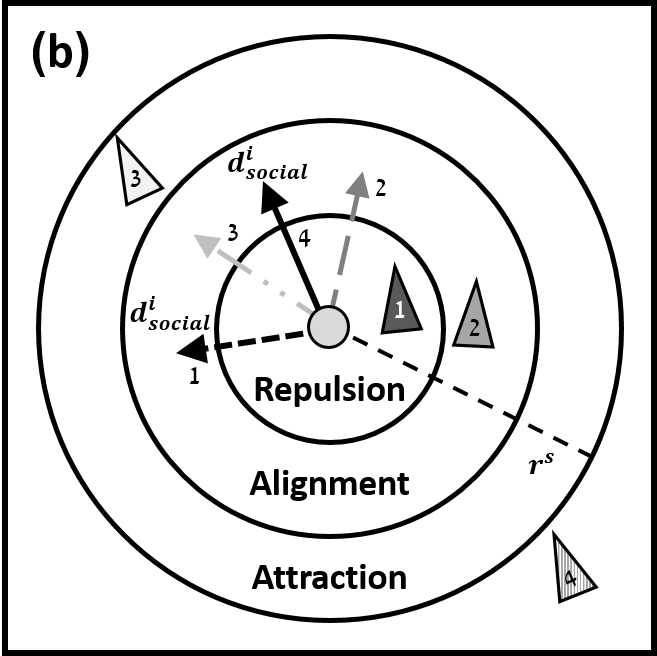}
        \captionsetup{font=scriptsize, labelfont={color=white}}
		\caption{}
    	\label{1b}
        \vspace{-0.5em}
	\end{subfigure}
	\begin{subfigure}{0.30\textwidth}
		\includegraphics[width=1\textwidth]{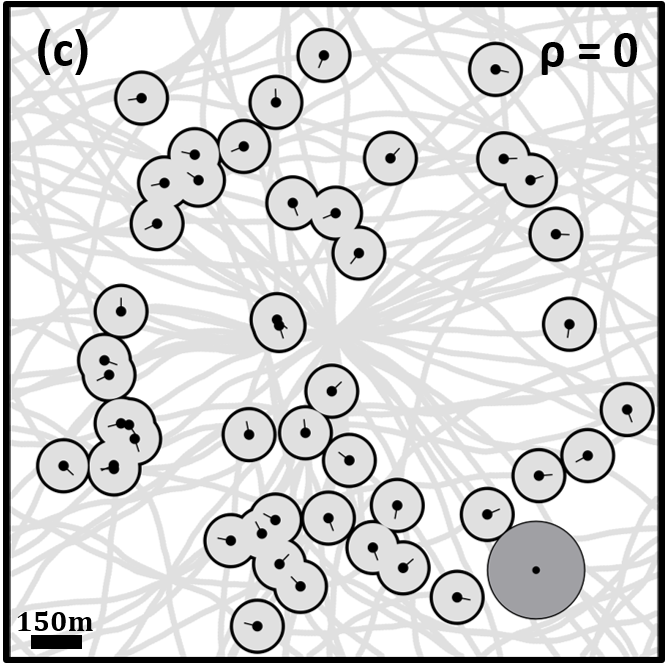}
		\captionsetup{font=scriptsize, labelfont={color=white}}
		\caption{}
        \label{1c}
        \vspace{-0.5em}
    \end{subfigure}\\
	\begin{subfigure}{0.30\textwidth}
		\includegraphics[width=1\textwidth]{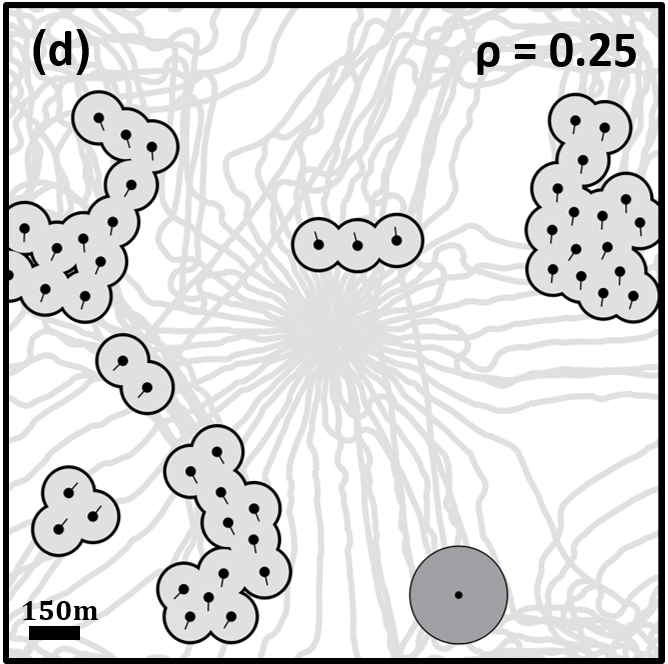}
	    \captionsetup{font=scriptsize, labelfont={color=white}}
		\caption{}
        \label{1d}
        \vspace{-1.5em}
    \end{subfigure}
	\begin{subfigure}{0.30\textwidth}
		\includegraphics[width=1\textwidth]{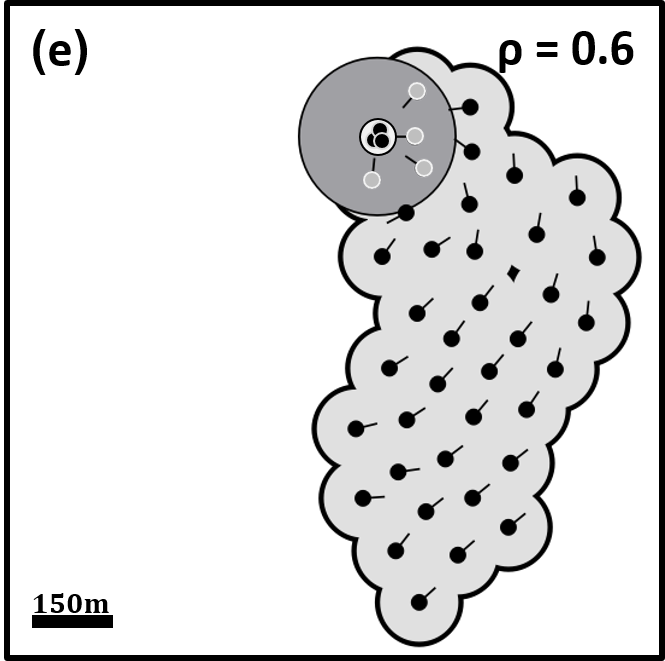} 
	    \captionsetup{font=scriptsize, labelfont={color=white}}
		\caption{}
        \label{1e}
        \vspace{-1.5em}
    \end{subfigure}
	\captionsetup{format=plain, labelfont=bf}
	\caption{The principles of the system. (a) The behavioral state diagram. $r^t$ is the target detection radius and $r^s$ is the social detection radius. When $r^s>r^t$ only the first agent will advance from the \emph{Search} state straight to the \emph{Find} state while the following agents will detect the first one and thus switch to the \emph{Lock} state. (b) The social interaction model includes three concentric zones: Attraction, Alignment and Repulsion. The dashed arrows represent the effect of each agent on the social direction of agent $i$. The black solid arrow is the average direction of arrows $2$ and $3$. $d^i_{\textrm{social}}$ is equal to arrow $1$ if the repulsion zone is non-empty (i.e. if agent $1$ exists) and equal to arrow $4$ otherwise. Note that the \emph{attraction} range is defined by $r^s$. Agents whose distance is larger than $r^s$ have no effect on $d^i_{\textrm{social}}$ (such as agent $4$). (c) A snapshot of the simulation where $\rho=0$. When $\rho$ was set to $0$, and hence individuals were moving independently, the only social interactions occurred when an agent homed in on a neighbor that already found the target (only when $r^s$ ~> $r^t$). The gray lines represent the agents' paths after $\sim400$ steps, the black circles represent the social detection region in which conspecifics can be sensed, the black lines (inside the circles) depict the heading of the agents and the dark-gray disc represents the food detection region. In panels c-e we simulated $n=50$ agents. In both (c) and (d) an agent has already detected the food and this is why the target's disc (dark gray) is defined by $r^s$. Note that the agents' circles have a radius of $\frac{r^s}{2}$ (and not $r^s$) because two agents must be at most $r^s$ apart in order to sense each other. In these simulations, the size of the search area was set to $4km^2$ instead of $400km^2$ for better visualization.(d) A snapshot of the simulation for the case where $\rho=0.25$. All symbols are the same as in c. Notice how groups are now formed. (e) A zoom-in on the target’s area demonstrates the convergence of the group on-to the target after one of the agents found it, for $\rho=0.6$. The white circle in the center depicts the detection range of the food with several individuals who found it. Individuals in the dark grey circle are locked on the individuals that have already found the food. All other individuals (the great majority) are moving in the direction of the food even though they have not detected it and even though they have not detected individual that already found food.} 
	\label{introduction} 
\end{figure*}

The social direction of agent $i$, $\ora{d_{\textrm{social}}^i}$, is calculated according to the direction and location of its neighbors. We adopt the commonly used 3-radii model \cite{Couzin2002} to govern the interactions between agents. In brief, each agent has three concentric zones in which: \emph{repulsion}, \emph{alignment} and \emph{attraction} govern the interactions respectively (Fig.~\ref{1b}). The agent will try to avoid collision with neighbors that are located in the \emph{repulsion} zone by moving away from them. If there are no neighbors in the \emph{repulsion} zone, the agent will align with neighbors that are located in the \emph{alignment} zone and it will be attracted to neighbors that are located in the \emph{attraction} zone. Note that the three interaction zones are limited by the social detection radius $r^s$. That is, only neighbors within a distance $r^s$ from the agent are sensed and interacted with. Essentially, the attraction zone was defined by $r^s$ while the two internal zones (\emph{repulsion} and \emph{alignment}) were smaller than $r^s$ and were determined experimentally (Methods). To simplify the model, we assumed that the weight of the contribution of each neighbor to $\ora{d^i_{social}}$, is equal and does not depend on its distance. 

The agents could be in one of three behavioral states: \emph{Search}, \emph{Lock} and \emph{Find} (Fig.~\ref{1a}). Initially, all the agents start at the \emph{Search} state. An agent that arrives within the sensing range of the target switches to the \emph{Find} state and moves directly to the target, while an agent that detects another agent in the \emph{Find} state switches to the \emph{Lock} state (we assume that individuals can sense when another individual within $r^s$ is feeding). An agent in the \emph{Lock} state homes in on the location of the agent that found the target. Importantly, agents in the search mode do not lock on agents in the lock mode. In other words, an agent only knows when another agent is feeding but not when another agent detected a feeding agent. All agents that arrive at the target remain there until the end of the simulation (we assume that food patches are very sparse but that they contain plenty of food so that there is no competition).Importantly, we also test the special condition where $r^s$=$r^t$. In this condition, agents can only be in the \emph{Search} or \emph{Find} states (but never in the \emph{Lock} state) and thus, agents only know the positions of their conspecifics, but not their state. This is because an agent will always find the food at the same instance as it finds an individual that found the food. This  is a very important condition, because if group foraging is beneficial under this condition, this implies that animals only need to know the positions of their neighbors and not their state in order to gain from group foraging, making it easier to evolve.  Figure~\ref{1c} and \ref{1d} and movies S1-2 demonstrate our model’s behavior for $\rho=0$ and $\rho=0.6$ respectively, we highly recommend viewing the movies. 

Because in nature mostly $r^t<r^s$, finding the target by the first agent results in an effective increase of the target-detection radius from $r^t$ to $r^s$, as from this point onward, the agents can detect the feeding individuals instead of the food itself (except for the special case described above where $r^t=r^s$). Such an increase occurs in nature, for example, when a bird of prey finds a carcass and circles it in the air or when a marine bird detects a school of fish and dives towards it repeatedly. Other birds can now home in on the carcass or the school from much larger distances by detecting the circling or the diving bird \cite{Harel2017, Jackson2008}. Similarly, when a bat finds a patch of insects, other bats can home in on this patch from larger distances by eavesdropping on the echolocation attack signals emitted by the first bat rather than by finding the insect patch \cite{Cvikel2015b}. 

Interestingly, our model generated a convergence effect on-to the food patch even though we did not explicitly define it. When one individual found the food and was moving towards it, other individuals followed this individuals attracting farther individuals long before they could directly sense the food or the finding individual. Therefore, the information about the location of food diffuses through the group attracting more and more individuals (from farther circles) long before they can sense the food itself or any of the individuals that already found it (Fig.~\ref{1e} and movie S1). This convergence occurred even in the special case where  $r^t=r^s$ in which individuals never sense that others have found food (they  only sense the food directly) proving that it is an emergent property of the movement. We next discuss the effect of different sensory and social model parameters on group searching performance. Importantly, we measured performance from an individual’s point of view as the average time to finding food.

\section*{Results}
To delineate the sensory regimes where grouping improves searching, we first set out to find the best non-social foraging model, that is, the model that minimizes the time for finding food when the individuals move without interacting. To this end, we set $\rho$ to $0$ and we varied $\sigma$ (the width of turning angle distribution) until we found the value that optimizes searching as individuals (minimized the mean searching time). The best $\sigma$ was small (in the range $0<\sigma \leq 3$ degrees, Fig.~S1) meaning that the agents moved almost in straight lines (we only allowed Gaussian noise in the turning angle distribution, so other movement distribution, such as Levy walks were not possible). Note  that we only varied $\sigma$ because in the non-social model, $\sigma$ is the only meaningful parameter while all other parameters are either irrelevant (because they describe interactions) or were set to be fixed in both the social and the non-social models (see Methods). In the process of finding the best non-social model (and everywhere else, unless stated otherwise), we simulated $50$ agents searching for a single target (food patch) in a $20km\cdot20km$ two-dimensional area with a detection ratio of 15 (specifically, we mostly used $\frac{r^s}{r^t}=\frac{150m}{10m}$). This ratio represents a common situation in which an animal can detect its prey from a shorter distance than it can detect its peers. This is, for example, the case for grazing sheep \cite{Ginelli2015} and for scavenging birds of prey \cite{Harel2017, Jackson2008}, and it is also the case for small birds that are searching for seeds on the ground and can see that another bird is pecking from much longer distances than they can detect the seeds themselves. The exact radius values that we started off with (i.e., $r^t=10m$ and $r^s=150m$) are typical for echolocating bats, which can detect prey from much shorter distance ($\sim10m$) than they can detect a neighbor ($\sim150m$). Bats can realize when a neighbor found prey based on its echolocation attack signals \cite{Cvikel2015b}. Below, we discuss the effect of varying this ratio and varying the specific radii. Note that even in the non-social model, individuals could recognize when an agent within $r^s$  from them has found the target and they moved towards it.

\begin{figure*}[h]
	\centering
	\begin{subfigure}{0.43\textwidth}
		\includegraphics[width=1\textwidth]{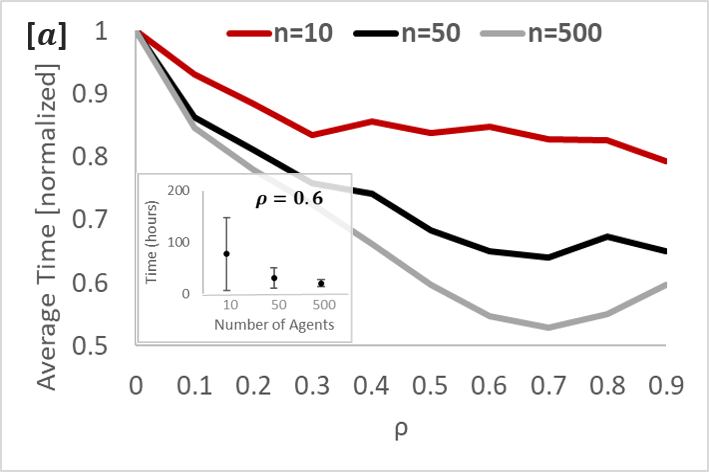} 
		\captionsetup{font=scriptsize, labelfont={color=white}}
		\caption{}
        \label{2a}
        \vspace{-1.5em}
	\end{subfigure}
	\begin{subfigure}{0.43\textwidth}
		\includegraphics[width=1\textwidth]{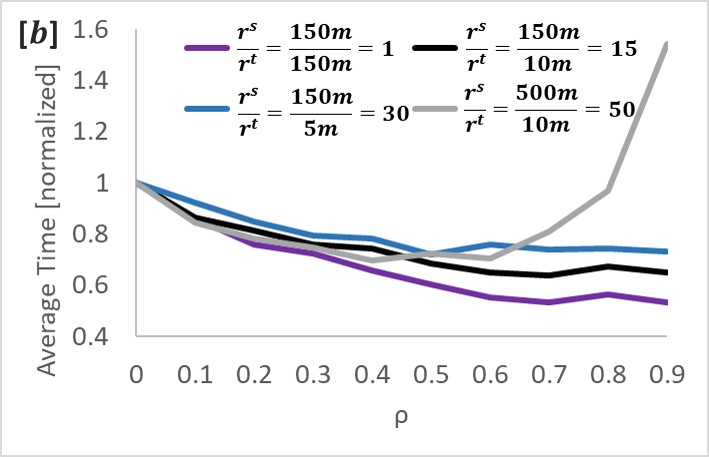} 
		\captionsetup{font=scriptsize, labelfont={color=white}}
		\caption{}
        \label{2b}
        \vspace{-1.5em}
	\end{subfigure}
	\begin{subfigure}{0.43\textwidth}
		\includegraphics[width=1\textwidth]{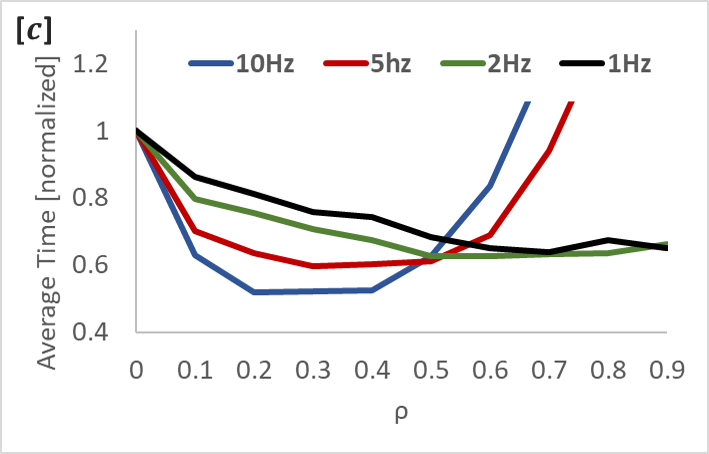} 
			\captionsetup{font=scriptsize, labelfont={color=white}}
		\caption{}
        \label{2c}
        \vspace{-1.5em}
	\end{subfigure}
	\begin{subfigure}{0.43\textwidth}
		\includegraphics[width=1\textwidth]{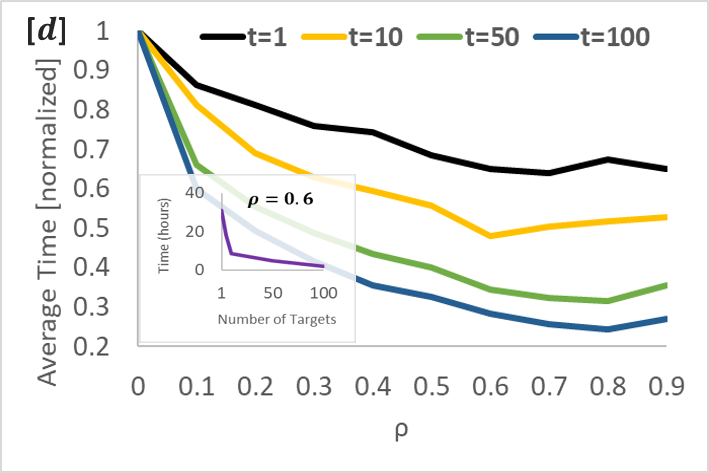} 
		\captionsetup{font=scriptsize, labelfont={color=white}}
		\caption{}
        \label{2d}
        \vspace{-1.5em}
	\end{subfigure}
	\captionsetup{format=plain, font=small, labelfont=bf}
	\caption{Social searching improves individual gain. The searching time in all panels was normalized by dividing the results by the searching time of the model where $\rho=0$ (each line in the graph was normalized separately) and the results are shown as an average over 1,000 simulations. The searching time was defined as the average time it took an agent to reach the target. In all panels, unless stated otherwise, the number of agents was 50, the target detection radius was set to $10m$ and the social detection radius to $150m$. In all panels, for each $\rho$ we plot the best result depending on $\sigma$ (deviation in the Gaussian noise of the individual direction). The black line is the same in all panels allowing comparison. (a) The mean searching time as a function of $\rho$ for different numbers of searching agents. The insert shows the results for $\rho=0.6$ without normalization. Note the reduction in both the mean searching time and the standard deviation. (b) The mean searching time depending on $\rho$ for different ratio between the social radius and the food detection radius. (c) The mean searching time as a function of $\rho$ for different values of $c_f$  (the movement direction update rate). (d) The mean searching time as a function of $\rho$ for different numbers of targets. Insert shows the results for $\rho=0.6$ without normalization to allow comparing the actual time to finding a target.} 
	\label{main_result} 
\end{figure*}

After finding the best non-social model, we used the same setup (e.g., same: area, number of agents and sensing radii), while varying the social weight $\rho$, simulating different degrees of social foraging (we tested the range: $0\leq\rho\leq0.9$). We assumed that finding a social model in which individuals perform better than in the best non-social model would demonstrate that social foraging is beneficial for searching. Indeed, increasing the social weight ($\rho$), improved searching by a factor of 1.6 in comparison to the best non-social model - the average time to find the food was 1.6 times faster (Fig.~\ref{2a}-\ref{2d}, there was a significant Pearson negative correlation between $\rho$ and the mean searching time, $P=0.005$. The difference between searching times at $\rho=0.6$ and $\rho=0$ was significant, $P<10^{-5}$, two-sample t-test). The best searching performance was observed when $\rho$ was between 0.6 and 0.9 (the exact value depended on the setup, see below). Social foraging improved searching in all group-sizes we examined (from 10 to 500 individuals), but it improved more in larger groups, reaching an improvement factor of 1.9 in a group of 500 agents (Fig.~\ref{2a}). Another advantage of searching in a larger group was a reduction in the searching time variability in accordance with the predictions of analytic models \cite{Clark1986, Pulliam1982}. The variance was estimated over multiple simulations and could be thought of as the variability in searching time over multiple days or nights of foraging. We took the overall variance of all individuals in all simulations with specific parameters (e.g. 50,000 individuals for $n=50$ individuals per simulation and $1,000$ simulations). The variability was smaller in larger groups (see insert in Fig.~\ref{2a} and Fig.~S2). Note that the agents did not necessarily form one large searching group that included all individuals, but that they typically split into multiple groups, which were disconnected from each other. In this analysis of group size, and in all further analyses, we always compared our social model to the best non-social model that had the same parameters. For example, if we changed the number of agents in the simulation; we also searched for the $\sigma$ that optimized the non-social model ($\rho=0$) with this new number of agents.

To further generalize our findings, we tested the effect of the sensing ratio ($r^s/r^t$) on searching in a group. Social searching was always better than the best non-social model, but the best social model (i.e., the best $\rho$) differed depending on this ratio (best $\rho$ was always between 0.5-0.9 Fig.~\ref{2b}). The advantage of social foraging decreased as the detection ratio increased, i.e., as the social detection range (of other individuals) increased relative to the target detection range ($r^s >> r^t$). The reason for this is that when individuals can detect other individuals from very large distances, they can home in on the food from very large distances once it is found by the first individual, so there is less need for searching in a group. Moreover, when ($r^s >> r^t$), moving independently actually has an advantage because the target is usually found faster by the first agent. This is why when the social detection range is very large (e.g., $r^s=500m$, grey line in Fig.~\ref{2b}), highly social models (large $\rho$ values) are detrimental; because large groups are formed and the target is found for the first time rather late. Notably, social searching was better than searching independently even in the special case in which the ratio was $1$ ($r^s= r^t$),  when agents never know when a neighbor finds food (because they arrive at the food and at the feeding individuals at the same time). 

Next, we tested how the optimal social model is influenced by another sensory parameter --- the sensory update rate --- the rate of updating the movement direction based on the positions of the neighbors ($c_f$, see Equation 5). So far, we assumed that agents update their movement direction once a second. Increasing this rate to 10Hz shifted the best $\rho$ from above 0.6 to around 0.3 (Fig.~\ref{2c}), and improved the performance of social foraging  even further in comparison with the non-social model up to factor of 1.9 (so far, the best 50-agent system achieved an improvement of 1.6-fold). The reason for this improvement was that larger and more spread-out groups were formed when a higher update rate was used (Movie S3). The range of the update-rates that we tested (1-10Hz) is typical for biological systems (e.g., \cite {Bar2015}).

Finally, we tested how the presence of multiple targets affects the results by introducing more targets in the area. We found that the advantage of group searching improved further when more than one target was present, reaching an improvement factor of 4.5 with 100 targets in the area (Fig.~\ref{2d}). This result proves that our findings hold also in the general case of many food sources.

\begin{figure*}
	\centering
	\begin{subfigure}{0.325\textwidth}
		\includegraphics[width=1\textwidth]{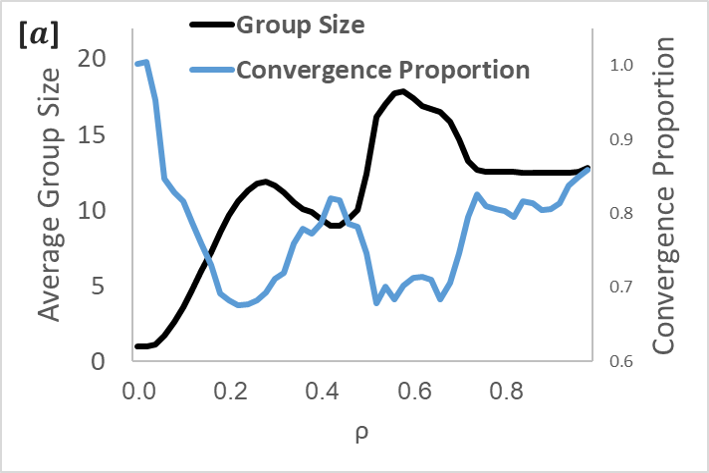} 
		\captionsetup{font=scriptsize, labelfont={color=white}}
		\caption{}
        \label{3a}
        \vspace{-1.5em}
	\end{subfigure}
	\begin{subfigure}{0.325\textwidth}
		\includegraphics[width=1\textwidth]{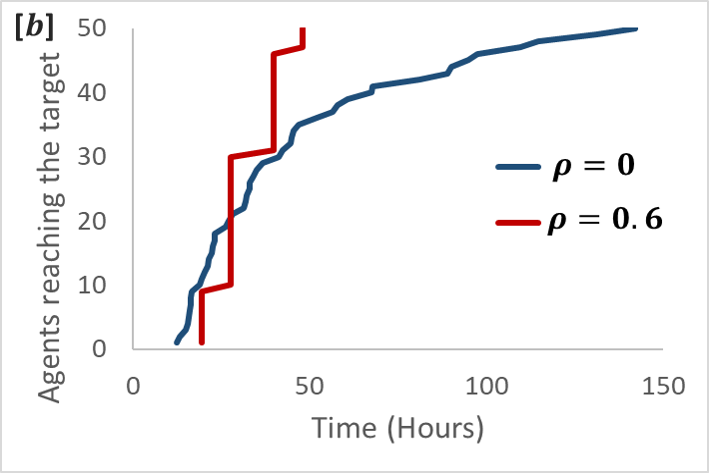} 
		\captionsetup{font=scriptsize, labelfont={color=white}}
		\caption{}
        \label{3b}
        \vspace{-1.5em}
	\end{subfigure}
	\begin{subfigure}{0.325\textwidth}
		\includegraphics[width=1\textwidth]{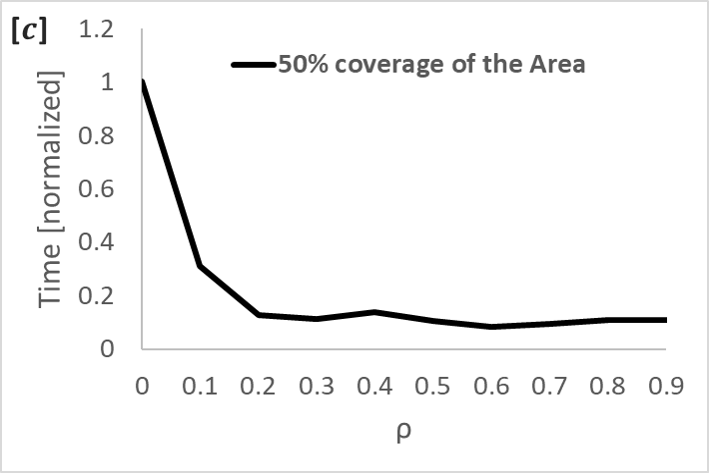} 
		\captionsetup{font=scriptsize, labelfont={color=white}}
		\caption{}
        \label{3c}
        \vspace{-1.5em}
	\end{subfigure}
	\captionsetup{format=plain, font=small, labelfont=bf}
	\caption{The mechanisms underlying improved social searching. In all panels, the number of agents was $50$, the food detection radius was set to $10m$ and the social detection radius to $150m$. Results are an average over $1,000$ simulations. (a) Group size and convergence proportion as a function of $\rho$. The convergence proportion was defined as the proportion of individuals that converged on-to the target when it was found by a member of the group. Note that the peaks in group size at $\rho=0.25$ and $\rho=0.6$ are not measurement errors. We ran this analysis with a resolution of 0.02 to confirm this. (b) Searching performance in a single simulation. The number of agents arriving at the target are shown as a function of time for $\rho=0$ (blue) and $0.6$ (red). When $\rho=0$, agents continuously reach the target as individuals. When $\rho=0.6$, agents reach the target in groups (as can be learned from the stair case shape of the graph). (c) The normalized cover time of $50\%$ of the area for the average group that is created at each $\rho$. For example, at $\rho=0.6$, the average groups size was $16$ agents, so we estimated the time it would take a group of $16$ agents to cover $50\%$ of the area. In both the Group Size line of panel a, and in panel c, we exclude the food from the searching area to examine searching dynamics without a reduction in the number of searching agents, which occurs when they reach the target (see Methods for more details).} 
	\label{explanation} 
\end{figure*}

An ideal searching system would be composed of independently moving agents with perfect communication between everyone, such that all individuals are informed when one finds food (assuming that competition over the food is not detrimental). Such a setup is non-realistic for most biological systems, and surely non-realistic for animals that must search large areas for food due to the limited range of social communication and eavesdropping (the range from which an individual can follow another one). In a realistic scenario, such as the one we modeled, two competing goals must be achieved for optimizing group searching. On the one hand, the group should spread as much as possible in order to minimize the searching overlap between its agents. On the other hand, individuals should maintain a connection to other group members so that they converge on-to the target when a member of the group finds it. Being part of a group does not help if the information about finding food by a group member does not reach other individuals. In practice, these two factors typically negatively correlate, meaning that a group that spreads out (i.e., with less searching-overlap) often has poor connection between its members. We found that the number of individuals per group increased rapidly with $\rho$, reaching a range of 10-16 individuals per group when $\rho$ was set to 0.2 or more (in a system with a total of 50 individuals). Larger groups did not form in our setup probably due to the set of interaction rules that we used. The increase in group size, was accompanied by a decrease in the proportion of converging individuals (Fig.~\ref{3a}). Convergence is less efficient in larger groups because each individual is influenced by more individuals some of which have not detected the food and have not detected individuals that already found the food. In other words, the movement is more noisy (large groups often split when arriving at the food). The best models showed a convergence proportion of 65-85\%, that is, when the target was found by a group member 65-85\% of the other members of the same group found it as well (see also Fig.~S3). Together, these effects determined the best social model (i.e., the best $\rho$).

The importance of the convergence effect for improved searching can be learned by observing the dynamics of finding food in a single simulation (Fig.~\ref{3b}). When $\rho$ is set to 0 (blue line) agents find food individually as can be learned from the monotonous increase of the proportion of agents who found the food. In comparison, when $\rho$ is set to 0.6 (red line), every event of finding of the food by an individual, is followed by many others converging on it as can be learned from the staircase shape of the graph. Note that when operating as individuals (blue line), the food is found faster for the first time (because individuals spread over the entire area rapidly), but very few other agents join (See Fig.~S4 for averages).

In addition to these two criteria (i.e., good convergence and little overlap), the group must also move efficiently in order to improve searching – a well spread group that hardly moves will not perform efficient searching. To quantify the overall searching performance of the model we estimated the time it takes groups of different size (equivalent to different values of $\rho$) to cover $50\%$ of the area (see Methods). This analysis (Fig.~\ref{3c}) shows that covering is better for social models ($\rho>0$). In fact, the best performance reaches a plateau from values of $\rho>0.2$ but because groups size is slightly smaller and so is the convergence rate (Fig. \ref{3a}) the overall best performance is achieved only at $\rho=0.6$ (for a system of $50$ agents, see also Fig.~S5).

\section*{Discussion}
By using a simple collective movement model we show that a group, with no leaders and without any direct communication between its individuals is more efficient in searching for food than individual agents. We tested a realistic biological model using biological plausible sensing ranges. We could have, for example, forced the agents to spread out in an optimally stretched line (with minimal overlap between them) and we could have imposed perfect communication between agents such that even when the most extreme agent in the line found food, all group members converge on it. This would have obviously improved searching, but such a behavior would require defining global rules. Instead, we used a model that only assumes simple local interactions between individuals. Still, even this simple model shows how improved searching can emerge spontaneously when animals move together.  Our general finding, that social searching improves searching was robust to various sensing and social model-parameters that we perturbed. We find that better group performance positively correlates with group size, with fast coverage of the areas and with good convergence once food is found (Fig. 3). However, we also show that there is a trade-off between group size and convergence --- as the group increases, less of its members will converge on-to the food when one of the members finds it. The actual improvement achieved by the group and the searching dynamics varied depending on the system's sensory, social and environmental parameters. It is likely that an organism that evolved to perform group searching, has evolved a different set of parameters and rules than those we used;  that would further improve searching. For example, such an organism could have evolved a higher sensory update rate, which we find advantageous under certain conditions. Additional sensory abilities that could be tuned by evolution to improve social foraging include an increase in the social detection range (of a conspecific) which could for example be achieved by improved sensing directionality or, in the case of sound emitting animals, by adapting the signal design. Importantly, these adaptations should result in a gain for the individuals in the group as our framework indeed suggests. In reality, the sensing radii and especially the food detection range $r^t$ might change seasonally and spatially depending on the available prey. Our prediction would thus be that the degree of social foraging within a species changes seasonally and spatially according to the season and region specific $r^t$ .

For group searching to improve searching, we only had to make two assumption: (1) Food is distributed in patches, that is, many group members can enjoy a food-patch when it is found (without competition), and (2) A neighbor’s location and movement direction can be tracked to some degree. Interestingly, it was not essential to assume that an agent can detect when another agent found the food. We originally assumed this, but found that it is not obligatory. This can be learned from the results of the model in the special case when $r^s=r^t$ (purple line in Fig.~\ref{2b}). When $r^s=r^t$ an individual must be within the target's detection range to sense that another individual found it, so it will have to reach the food itself and will never detect an individual that has already found the food. Even under this condition, the social model performed better than the non-social model. This condition also clarifies that the detection range of another agent ($r^s$) must not be larger than the target's detection range ($r^t$) for social foraging to be beneficial. Our original intuition was that if the two sensory ranges are the same (as in this case) there is no benefit to the group, but our results show otherwise. The reason that group foraging is beneficial even in this condition, is the convergence effect, which occurs when someone finds the target and occurred even when $r^s=r^t$. Namely, even if the target is only sensed when the agent reaches it (i.e., is within $r^t$ from it), when a group member reaches the target, its movement will attract others in its direction, leading to a convergence on-to the target. 
In fact, social searching was more important in the situation when $r^s=r^t$ than when $r^s>>r^t$  (compare purple and gray lines in Fig.~\ref{2b}). In contrast,  and slightly counter-intuitive at first, the only situation where the benefit of social foraging was very small, based on our model, was when $r^s$ was very large, that is when individuals can sense the location and state of other individuals from a large distance (e.g., grey line in fig. 2b). In this situation, agents can home in on a food item that was found by another individual from a very large distance, so there is less advantage in searching together. This might be the case for vultures that can detect individuals circling a carcass from very large distances.

Importantly, the basic interaction model that we used, was sufficient to generate a diffusion of information about the location of the target and a convergence on-to it once it was found by one of the group members.  We did not have to explicitly model this behavior, that is, the same interaction rules that were applied during search also resulted in this convergence.  This can be learned, once again, from the special case where $r^s=r^t$ because in this condition individuals never get the chance to know when another individual found food (if they are close enough to a feeding individual, they will already detect the food itself). Moreover, our results suggest that the internal zone of the 3-radii model --- the repulsion zone --- which has been originally designed to prevent collisions between agents, also assists social searching, because it forces the group to spread out and cover more area while searching.

As much as we increased $\rho$, the typical group size did not exceed $17$ individuals (larger groups split into several smaller groups). When we tuned other model parameters to force the formation of larger groups, their movement was often not efficient and they tended to get stuck in place. Indeed, in nature, foraging animals are often seen in groups of several individuals to dozens of individuals while swarms of thousands or more (which are often mathematically modeled) are more commonly observed in sleeping sites or in stationary situations when the group remains in place aiming to avoid predation \cite{Hoare2004}. Mathematical models of group size (e.g., \cite{Pulliam1984}) rarely consider group-movement constraint which seem to play a role based on our model. But, even though the size of a typical group did not grow, when there were more agents in the area, searching became more efficient for the average individual.  Food was found faster and the inter-simulation variability decreased dramatically (see insert in Fig.~\ref{2a}). This was mainly a result of the increase in the target's detection range from $r^t$ to $r^s$ once it was found by the first agent. This improvement with number of agents was thus not relevant for systems in which $r^s=r^t$. 

All of the system parameters that we tested influenced the performance of the model, but in all cases, there was always a social model that was better than the best comparable individual model. In reality, animals that evolved to rely on social searching might be flexible, adjusting how they weigh personal and social information according to the specific scenario, i.e., the type of food that is available, the sensory modality they are using to search for it and the sensory modality they are using to follow neighbors. Our agent based model thus suggests how we can use simulations to gain intuition about social foraging and more generally, how we can gain insight about the reasons underlying field observations on animal behavior. 

\section*{Methods}
\textbf{Experimental setup.} The model was defined by a set of environmental parameters and agent parameters. The environmental parameters included: $S$ – the size of the searching area, $n$ – the number of agents and $t$ – the number of targets (whose positions were sampled from a uniform distribution at each simulation). The agent parameters included: $v$ – the agent’s velocity, $c_f$  - the movement direction update rate, $r_t$ – the target detection range, $r_s$ – the social detection range, $\rho$ - the weight of the neighbors’ influence on the agent’s direction, $\sigma$ - the width of the distribution of turning angles of the Self Direction and $r_1$, $r_2$ - the radii of the concentric (repulsion and alignment) zones (where $r_3=r_s$). We refer to a model with a specific set of parameters as an instance of the model.
In our experiments, we fixed some of the parameters based on a simplified but realistic scenario. The fixed parameters were: $v = 10m/s$ and $S=20km\cdot20km$. We simulated a two dimensional world with toroidal boundaries, meaning that when an agent reached a border of the searching area it reappeared on the opposite side. This methodology is commonly used \cite{Giuggioli2016FromAT} to avoid the edge case when the agents reach the borders. In most simulations we also fixed the following parameters: $n=50$, $t=1$ ,$c_f=1Hz$, $r^t=10m$ and $r^s=150m$, but unlike $v$ and $S$, these parameters were varied in specific analyses to test their influence on performance . The main aim of this study was to find out whether social behavior improves searching. To that end, we started our experiments by searching for the best instance of the model where the agents search as individuals, i.e. $\rho=0$ (see below the definition of searching performance). Note that when $\rho=0$ the model depends only on the distribution of turning angles, determined by $\sigma$ ($r_1$ and $r_2$ are meaningless in this case). We therefore searched over different values of $\sigma=\{0-10, 15, 20, 30, 50\}$ running $1,000$ simulation for each instance. Then we searched for a social model where $\rho>0$ improves searching performance. In order to reduce the number of model parameters we first searched for concentric radii values that showed good performance. We did this by searching over four parameters simultaneously: $\rho=\{0.1 : 0.1 : 0.9\}$, $\sigma=\{1, 3, 5, 10, 15, 20, 30, 50\}$, and $r_1=\{0 : 0.1*r^s : r^s\}$ and $r_2=\{0 : 0.1*r^s : r^s\}$ ($r_1 \leq r_2$), total of $\sim4,500$ instances (4-parameters sets); each instance was tested over $50$ simulations. According to the results, we fixed the concentric radii, $r_1$ and $r_2$ that achieved on average the minimal searching time: $r_1$ was set to $\frac{3}{10}r^s$ and r2 was set to $\frac{7}{10}r^s$. Later, we isolated some of the additional agent parameters and tested their effect on the performance of the model while focusing on changing the social weight $\rho=\{0 : 0.1 : 0.9\}$ (we always tested different values of $\sigma=\{3, 10, 20, 30, 50\}$ for each model instance Fig.~S6). In these experiments (after $r^1$ and $r^2$ were fixed) each instance was tested over $1,000$ simulations. We tested the model for different numbers of agents, $n=\{2, 5, 10, 50, 100, 200, 500\}$, different ratios between the social detection radius to the target detection radius, $\frac{r^s}{r^t}=\{1, 15, 100\}$, different rates of updating the movement direction, $c_f=\{10Hz, 5Hz, 2Hz, 1Hz\}$ and different numbers of targets, $t=\{1, 2, 5, 10, 30, 50, 100\}$.

\textbf{Metrics.} In order to evaluate the model and explain its behavior we defined the following metrics (see below for formal definitions): $S_{\textrm{avg}}$ - the mean searching time to find a target (Fig.~\ref{2a}-\ref{2d}); $C_{\textrm{prop}}$ - the convergences proportion of a subgroup on-to a target (Fig.~\ref{3a});  $G_{\textrm{size}}$ - the average group size (Fig.~\ref{3a}); $f_{\textrm{ac}}^k(x)$ - the time it takes a subgroup of $k$ agents to cover $x$ percent of the searching area (Fig.~\ref{3c}). Note that in order to estimate the last three metrics, we excluded all targets from the search area and ran the simulation for a constant time of $500K-5,000K$ time steps (depending on $c_f$). We used this approach when we did not want to be affected by the convergence of individuals on-to the target, which decreased the number of individuals over time. The metrics were evaluated in each simulation and the reported results are average across all simulations. 

Next, we formally define the metrics. The mean searching time to find a target, $S_{\textrm{avg}}$, is defined as follows,
\begin{align}
	S_{\textrm{avg}} = \frac{1}{n}\sum\limits_{i=1}^{n}{s_i},
\end{align}
where $n$ is the number of agents and $s_i$ is the time it took agent $i$ to find the target.

To define $C_{\textrm{prop}}$ and $f_{\textrm{ac}}^k(x)$ we will first define the average number of connected components, $C_{\textrm{comp}}^*$,
\begin{align}
	C_{\textrm{comp}}^* = \frac{1}{T}\sum\limits_{t=0}^{T}{C_{\textrm{comp}}(G^t)},
\end{align}
where $n$ is the number of agents, $T$ is the number of iterations of the simulation, $C_{\textrm{comp}}(G^t)$ is the number of connected components in the graph $G^t$. The Graph $G^t$, is defined as follows,
\begin{align}
	G^t(V,E^t)=
	\begin{cases}
		V = \{v_1,...,v_n\}\\
         \!\begin{aligned}
            & E^t = \{<v_i,v_j>\}, \\
            & <v_i,v_j>\in E^t \ \rm{iff}
        	\norm{p_i^t-p_j^t}_2 \leq r^s,
         \end{aligned} 
	\end{cases}
\end{align}
where $n$ is the number of agents, $p_i^t$ is the position of an agent $i$ at time $t$ and $r^s$ is the social radius. We match a graph for each time step. The vertices of the graph represent the agents, that is $v_i$ represent agent $i$. The edges of the graph represent which agents are in range of detection from each other. That is, we connect two vertices by an edge if the distance between their corresponding agents is less or equal to $r^s$ (the social detection radius). For each such graph we calculated the number of connected components and $C_{\textrm{comp}}^*$ is the average number of connected components over all of the graphs in a single simulation.

The convergence proportion of a subgroup to a target, $C_{\textrm{prop}}$, is an estimation of the number of agents that will reach the target immediately after it is found by one of the subgroup members (it is possible that some of the agents will miss the target). $C_{\textrm{prop}}$ is defined as follows,
\begin{align}
	C_{\textrm{prop}} = \frac{C_{\textrm{comp}}^*}{F_{\#}},
\end{align}
where $C_{\textrm{comp}}^*$ is the average number of connected components and $F_{\#}$ is the number of times that the target is found directly (i.e., not because an agent followed another agent to the target). As $F_{\#}$ is closer to $C_{\textrm{comp}}^*$ the convergence effect is larger, i.e. more agents of a subgroup will reach the target when it is found by one of the subgroup's agents.
$F_{\#}$ is defined as follow,
\begin{align}
	F_{\#} = \sum\limits_{i=1}^{n}{x_i}
\end{align}
where $x_i$ indicates (with high probability) if agent $i$ found the target by himself, i.e. not because it followed another agent. $x_i$ is defined as follow,
\begin{align}
	x_i = 
	\begin{cases}
		0, \exists{j\neq{i}}\textrm{ s.t. }0<t_i-t_j\leq\frac{r^s*c_f}{v},\\
		1, \textrm{otherwise},
	\end{cases}
\end{align} 
where $t_i$ is the searching time of agent $i$, $r^s$ is the social detection radius, $c_f$ is the movement direction update rate and $v$ is the agent's velocity. Agents can effect each other only if the distance between them is less than $r_s$. At each step an agent advances $\frac{v}{c_f}$, which means that it will take an agent $\frac{r^s*c_f}{v}$ time steps to advance a distance of $r_s$. If the reaching time between agent $i$ and all the other agents (that reach the target before him) is greater than $\frac{r^s*c_f}{v}$ then we can conclude that agent $i$ found the target by himself. On the other hand, if there is an agent $j$, that reached the target before agent $i$ and $t_i-t_{j}\leq\frac{r^s*c_f}{v}$ we can conclude with high probability that agent $i$ followed agent $j$ to the target. We can not be sure that agent $i$ followed agent $j$ because it is possible that they reached the target from different directions.


The average group size, $G_{\textrm{size}}$, is defined as follows,
\begin{align}
	G_{\textrm{size}} = \frac{n}{C_{\textrm{comp}}^*}
\end{align}
where $n$ is the number of agents and $C_{\textrm{comp}}^*$ is the average number of connected components.

To estimate the value of $f_{\textrm{ac}}^k(x)$, namely the time it takes a subgroup of $k$ agents to cover $x$ percent of the searching area, ,we first calculated the accumulated cover time of the complete group, $f_{\textrm{ac}}^n(x)$. To do so, we divided $S$ in to an $m\times m$ grid, each cell of size $\frac{r^t}{2}\times \frac{r^t}{2}$. Each cell that is visited by an agent is considered covered. Once a cell is covered, it stays that way until the end of the simulation. $f_{\textrm{ac}}^n(x)$ is calculated as follows,
\begin{align}\label{cover time}
	f_{\textrm{ac}}^n(x) \sim \textrm{min}\{t\geq0 : \frac{m_t}{M}\geq \frac{x}{S}\},  
\end{align}
where $m_t$ is the number of covered cells at time $t$ and $M$ is the total number of cells in the grid. $f_{\textrm{ac}}^k(x)$ is calculated as follows, 
\begin{align}
	f_{\textrm{ac}}^k(x) \sim \frac{n}{G_{\textrm{size}}}f_{\textrm{ac}}^n(x),  
\end{align}
where $n$ is the number of agents, $G_{\textrm{size}}$ is the average number of agents in a subgroup and $f_{\textrm{ac}}^n(x)$ is the accumulated cover time of the complete group. To estimate $f_{\textrm{ac}}^k(x)$ we measure the accumulated cover time of the complete group and then we normalized it by the number of agents divided by the average group size. Another approach to measure $f_{\textrm{ac}}^k(x)$ is to run a simulation with $k$ agents, $1\leq k\leq n$. In addition to requiring more simulations this latter approach is also not accurate. This is because a simulation with a group of $k$ agents will split into subgroups which will reduce the overlap between agents and increase the cover rate efficiency.

\paragraph{ACKNOWLEDGMENTS.} We thank H. Balaban and N Gov for commenting on the manuscript. This project was partially funded by ONRG grant no. N62909-13-1-N066. Work by D.H.\ and R.C.\ has been supported in part by the Israel Science Foundation (grant no.~825/15), by the Blavatnik Computer Science Research Fund,
by the Blavatnik Interdisciplinary Cyber Research Center at Tel Aviv
University, and by grants from Yandex and from Facebook.



\bibliographystyle{unsrt}  
\bibliography{ref}

\newpage
\onecolumn
\setcounter{figure}{0}
\section*{\Huge Supplementary}

\begin{figure}[h]
\vspace{5cm}
\centering
\includegraphics[width=0.9\textwidth]{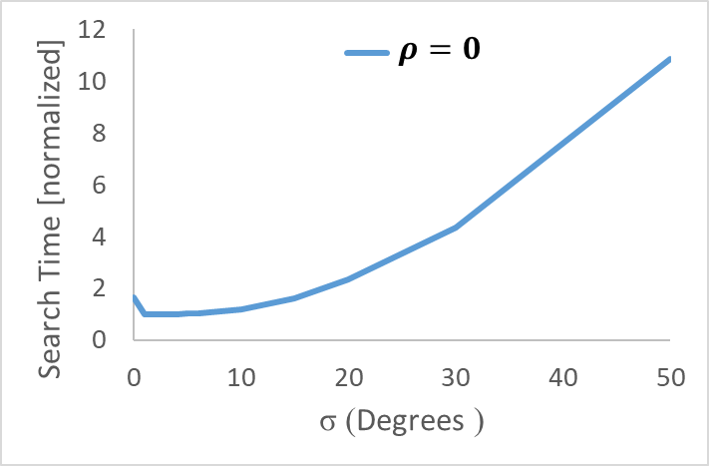}
\captionsetup{width=.9\linewidth, format=plain, font=small, labelfont=bf}
\caption{The mean searching time for the non-social model as a function of the width of the turning angle distribution ($\sigma$).}
\end{figure}

\newpage

\begin{figure}[h]
\vspace{5cm}
\centering
\includegraphics[width=0.9\textwidth]{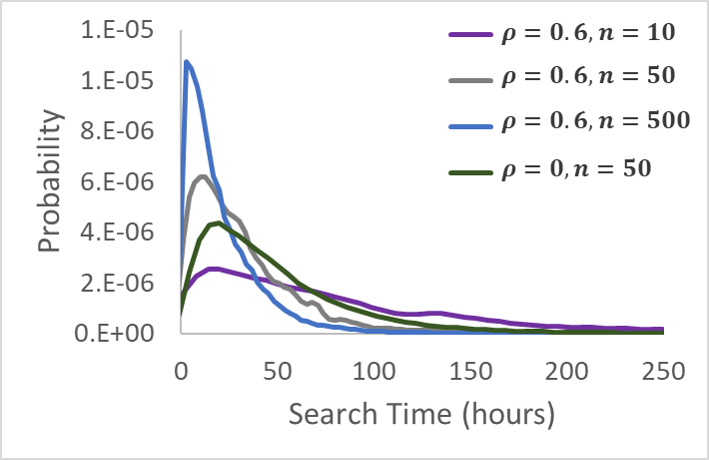}
\captionsetup{width=.9\linewidth, format=plain, font=small, labelfont=bf}
\caption{The probability density function of the mean searching time for different values of $\rho$ and $n$. Note how the mean of the distribution decreases for $\rho>0$ and how the width decreases with $n$.}
\end{figure}

\newpage

\begin{figure}[h]
\vspace{5cm}
\centering
\includegraphics[width=0.9\textwidth]{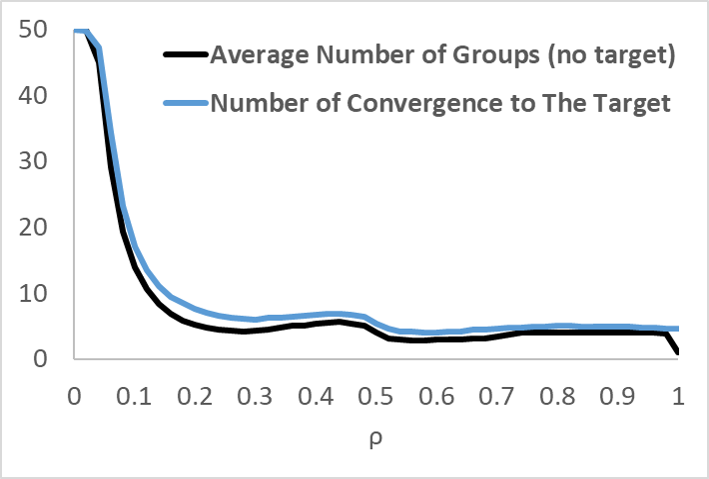}
\captionsetup{width=.9\linewidth, format=plain, font=small, labelfont=bf}
\caption{The number of groups and the number of convergence events as a function of $\rho$. When the two numbers are the same this implies that all group members converged when one of them found the target. The larger the difference between the two - the poorer the convergence.}
\end{figure}

\newpage

\begin{figure}[h]
\vspace{5cm}
\centering
\includegraphics[width=0.9\textwidth]{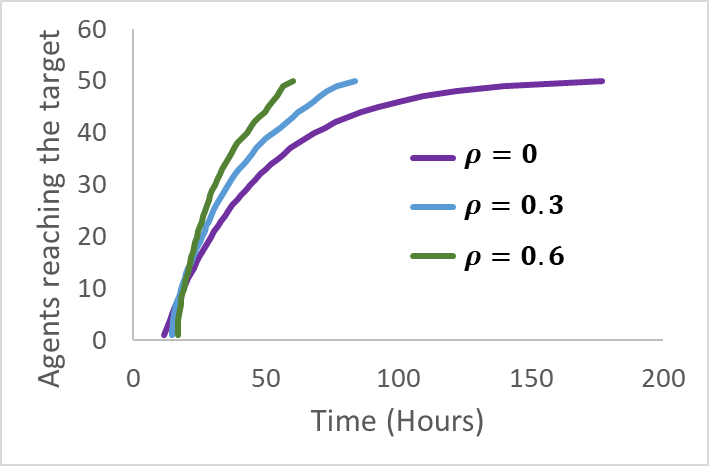}
\captionsetup{width=.9\linewidth, format=plain, font=small, labelfont=bf}
\caption{Number of agents reaching the target as a function of time for different values of $\rho$. For each $\rho$ we present the results of the best $\sigma$. The social graphs ($\rho>0$)  do not show steps (like in Fig. 3b) because they are an average of $1,000$ simulations. In these simulation the following parameters were used: $n=50$, $\frac{r^s}{r^t}=15$, $\sigma=3$, $t =1$.}
\end{figure}

\newpage

\begin{figure}[h]
    \vspace{1cm}
	\centering
	\begin{subfigure}{0.45\textwidth}
		\includegraphics[width=1\textwidth]{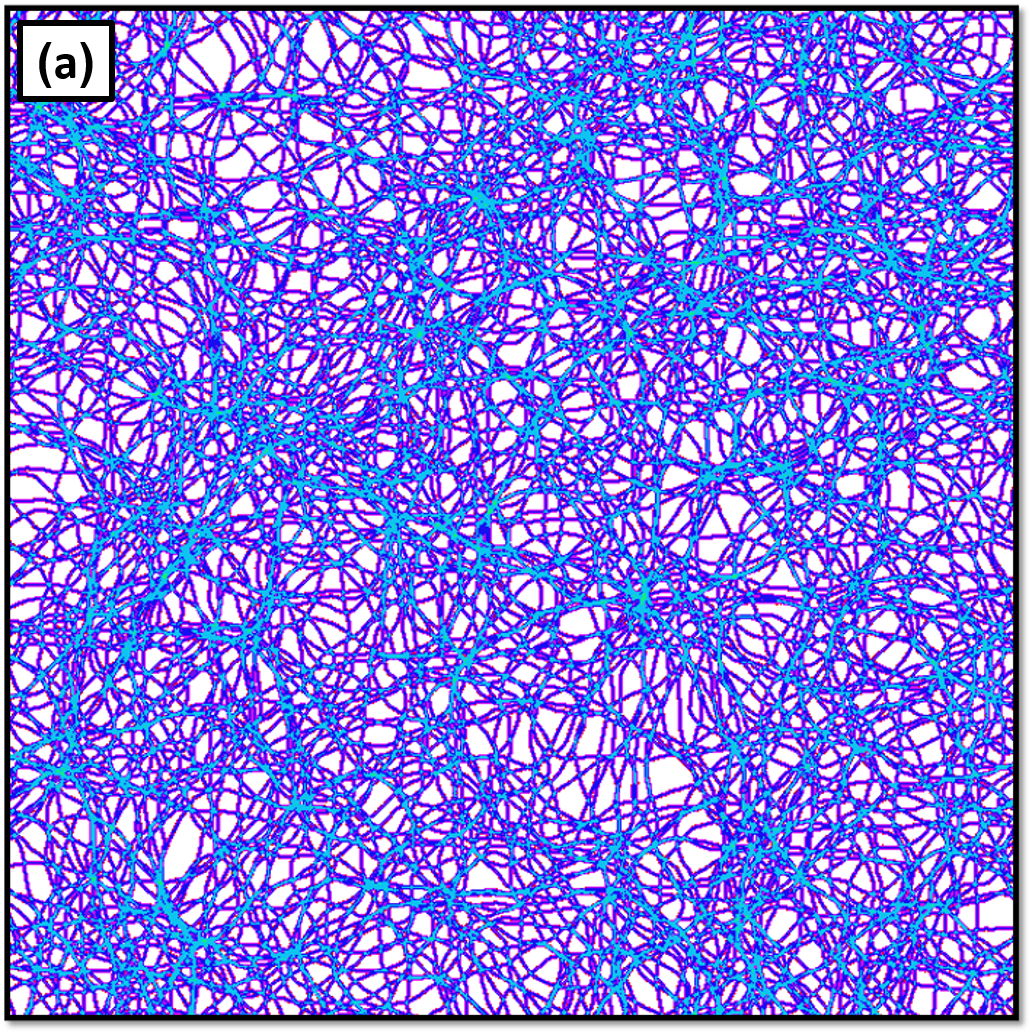} 
	\end{subfigure}
	\begin{subfigure}{0.45\textwidth}
		\includegraphics[width=1\textwidth]{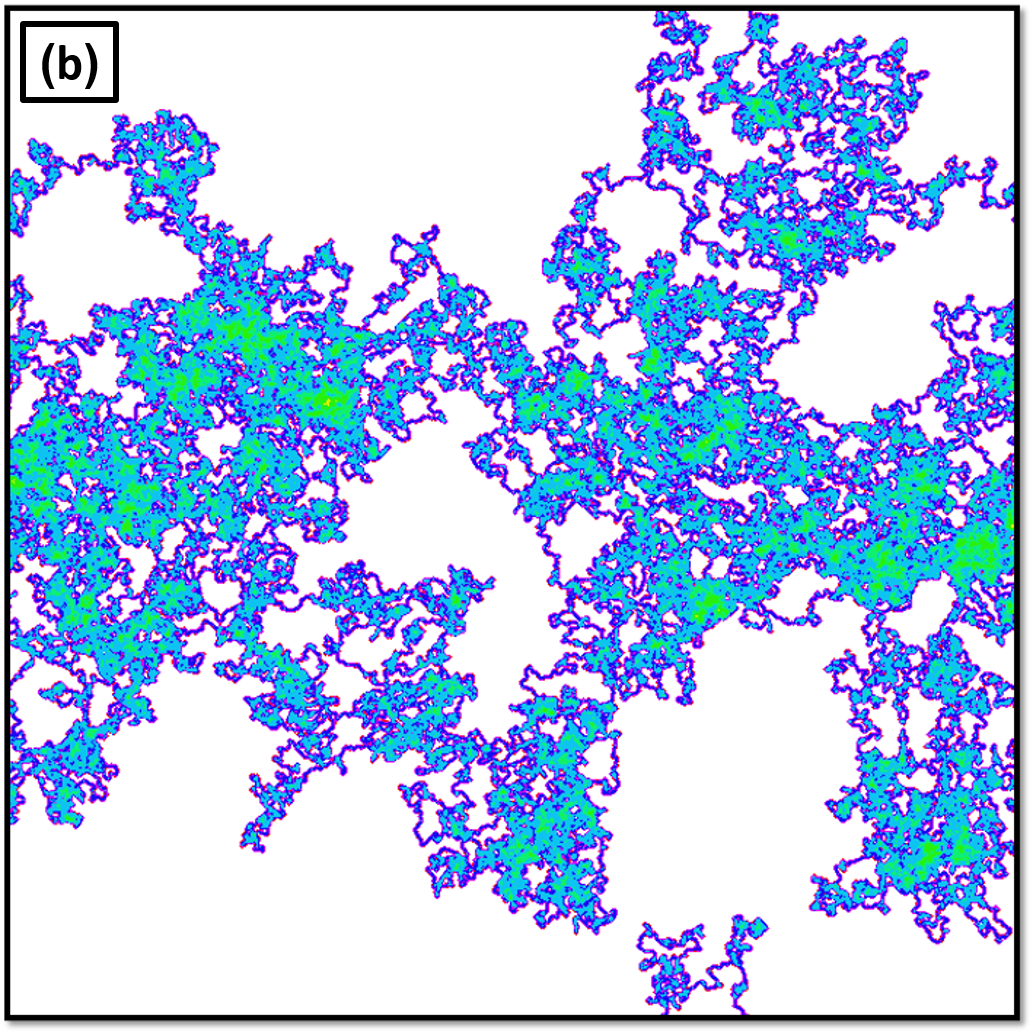} 
	\end{subfigure}
	\begin{subfigure}{0.45\textwidth}
		\includegraphics[width=1\textwidth]{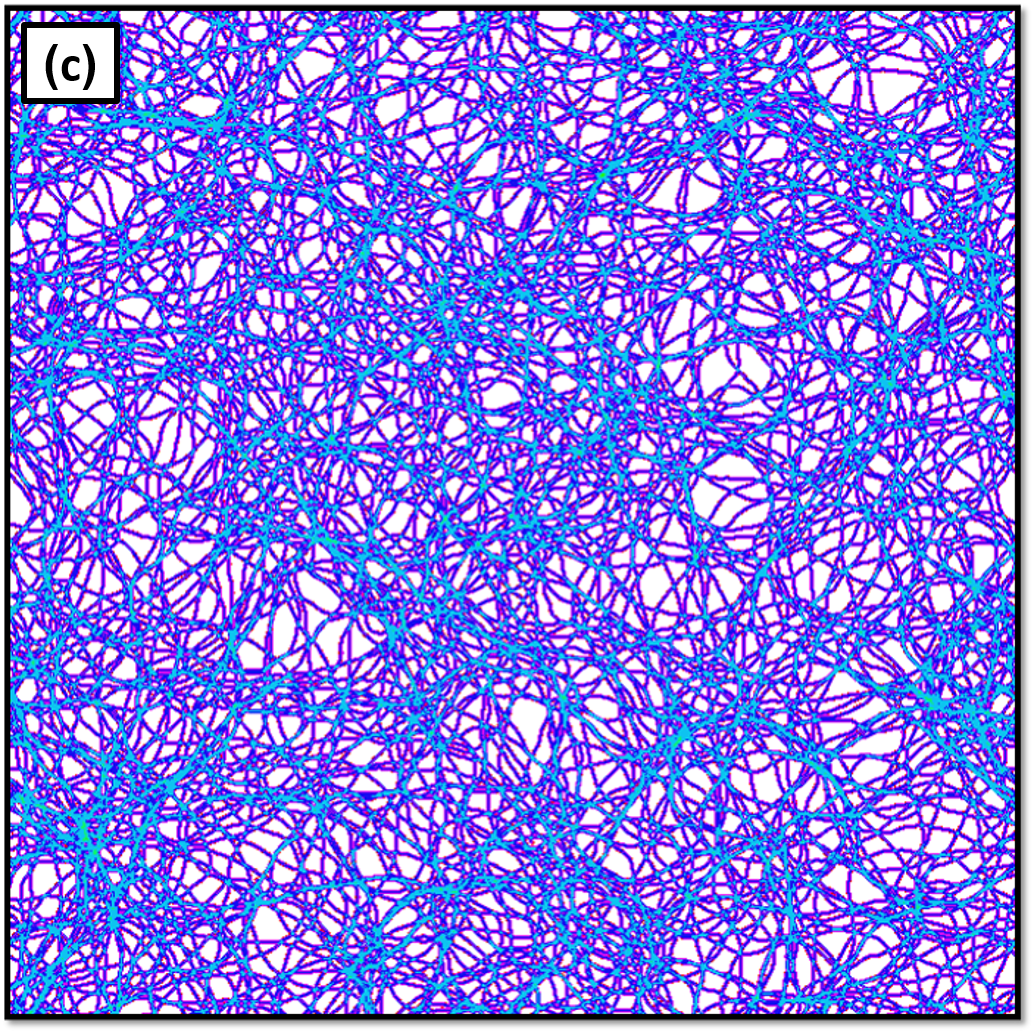} 
	\end{subfigure}
	\begin{subfigure}{0.45\textwidth}
		\includegraphics[width=1\textwidth]{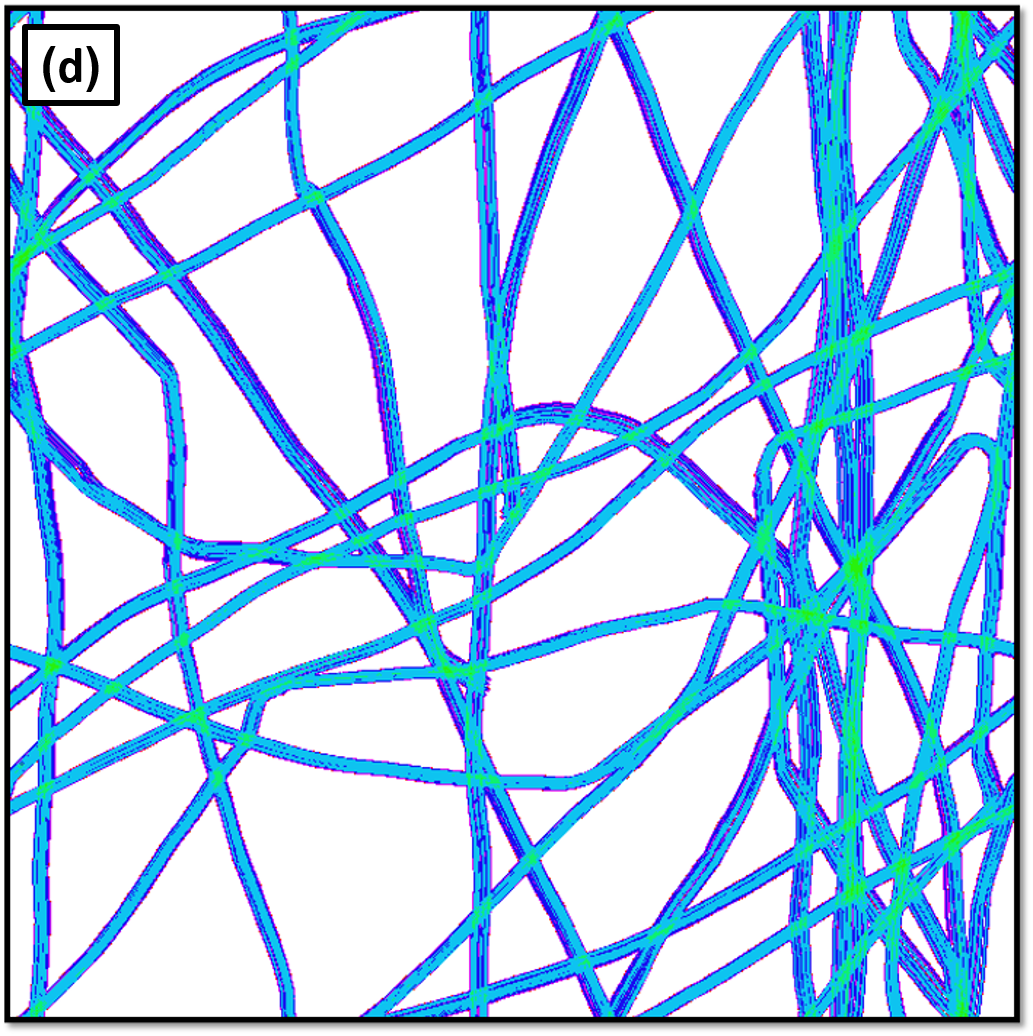} 
	\end{subfigure}
	\captionsetup{width=.9\linewidth, format=plain, font=small, labelfont=bf}
	\caption{Snapshots of the covered area for different models. We exclude the target from the area. (a) $n=1$, $r^t=30m$, $\rho=0$, $\sigma=3$, $T=500,000$ time steps ($\sim140$ hours). (b) same as a. but with $\sigma=30$. (c) $n=10$, $r^t=30m$, $\rho=0$, $\sigma=3$, $T=50,000$ ($\sim14$ hours). (d) same as c. but with $\rho=0.6$.} 
	\label{} 
\end{figure}

\newpage

\begin{figure}[h]
\vspace{5cm}
\centering
\includegraphics[width=0.9\textwidth]{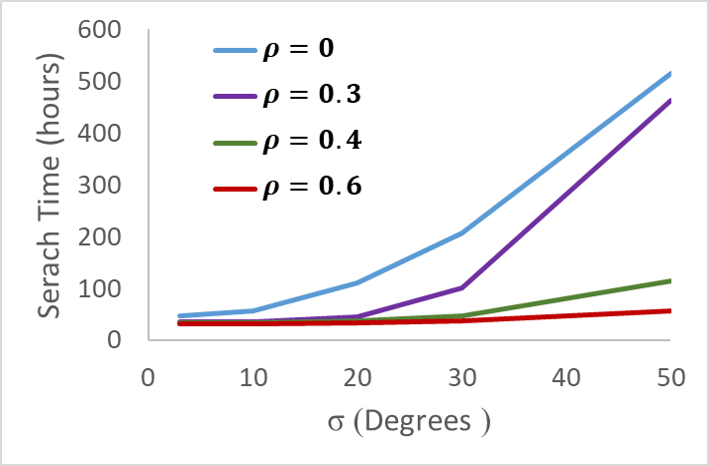}
\captionsetup{width=.9\linewidth, format=plain, font=small, labelfont=bf}
\caption{Mean searching time for different combinations of $\rho$ and $\sigma$. In these simulation the following parameters were used: $n=50$, $r^s/r^t =15$,  $t=1$.}
\end{figure}

\newpage

\begin{flushleft}
\textbf{Movie $S_1$.} \href{https://mailtauacil-my.sharepoint.com/:v:/g/personal/ravidc_mail_tau_ac_il/Eczr-p3lLDJPmp7JTRhuAPAB-KdD_UaoYMDqh5Pky3wI4g?e=rKlAY7}{Demonstration of individual search}.\\
\vspace{5mm}
\textbf{Movie $S_2$.} \href{https://mailtauacil-my.sharepoint.com/:v:/g/personal/ravidc_mail_tau_ac_il/EW3b0JaTbsNLl_6fX0gX_pABPJ2CgFwbNLiO6bFyC_GOKw?e=3E4h56
}{Demonstration of social search}.\\
\vspace{5mm}
\textbf{Movie $S_3$.} \href{https://mailtauacil-my.sharepoint.com/:v:/g/personal/ravidc_mail_tau_ac_il/Eecj8zuxz1tPlxIU7QuKz9sBBCBJDaHZDwwWmqij5_KzDg?e=CasJnZ
}{Demonstration of group motion for different $\rho$ and $\sigma$}.
\end{flushleft}

\end{document}